\documentclass[10pt,preprint2]{aastex}

\slugcomment{manuscript for AJ, version 1.6 170215}

\shorttitle{UCAC5 proper motions}
\shortauthors{Zacharias}

\begin{document}


\title{UCAC5: New Proper Motions using Gaia DR1}


\author{N. Zacharias\altaffilmark{1}, C.~Finch\altaffilmark{1},
   J.~Frouard\altaffilmark{1} }
\email{nz@usno.navy.mil}

\altaffiltext{1}{U.S.~Naval Observatory, Washington DC}


\begin{abstract}
New astrometric reductions of the US Naval Observatory CCD Astrograph
Catalog (UCAC) all-sky observations were performed from first
principles using the TGAS stars in the 8 to 11 magnitude 
range as reference star catalog.  Significant improvements in the 
astrometric solutions were obtained and the UCAC5 catalog of mean
positions at a mean epoch near 2001 was generated.
By combining UCAC5 with Gaia DR1 data new proper motions on the Gaia coordinate
system for over 107 million stars were obtained with typical accuracies of 
1 to 2 mas/yr (R = 11 to 15 mag), and about 5 mas/yr at 16th mag.
Proper motions of most TGAS stars are improved over their Gaia data
and the precision level of TGAS proper motions is extended to many
millions more, fainter stars.
External comparisons were made using stellar cluster fields and
extragalactic sources.
The TGAS data allow us to derive the limiting precision of the UCAC
$x,y$ data, which is significantly better than 1/100 pixel.
\end{abstract}



\section{Introduction}

The very successful European Space Agency (ESA) Gaia mission is
in progress.  In September 2016 the first Gaia data were released 
based on the first 14 months of regular in-orbit operations \citep{G1}.
Accurate mean observed positions for the 2015.0 epoch were provided
for a total of over 1.1 billion stars down to about magnitude G = 20.7 
with a median position error of about 1.8 and 1.6 mas for RA and Dec
respectively \citep{G4}.
A full 5-parameter (position, proper motion, parallax) astrometric
solution was obtained for just over 2 millions of stars, a subset of
about 80\% of the Hipparcos and Tycho-2 stars, by using the Hippacos
satellite observations of mean epoch 1991.25 together with the Gaia 
observations to resolve the proper motion and parallax degeneracy.
The accuracy of these positions varies as function of magnitude and
location on the sky with a median near 0.3 mas per coordinate
and a median error in proper motion of 1.1 and 0.9 mas/yr for RA and
Dec, respectively \citep{G4}.

The U.S.~Naval Observatory (USNO) is engaged in producing astrometric
star catalogs.  The USNO CCD Astrograph Catalog (UCAC) project provided
such all-sky data to 16th magnitude with its most recent 4th data release,
the UCAC4 \citep{U4}.  The Tycho-2 catalog \citep{T2} was used as
reference star catalog for those wide-field CCD observations.
Comparisons between the Tycho-Gaia Astrometric Solution (TGAS) and 
Tycho-2 proper motions showed that
the latter have large sky-correlated systematic errors of up to a
few mas/yr \citep{G4} due to previously undiscovered large systematic
errors in the early epoch Astrographic Catalog data from around 1900,
which were used for Tycho-2 proper motions.
A re-reduction of the UCAC data using TGAS as described in this paper
provides a significantly improved product, the UCAC5.

Although the UCAC has positional accuracies not reaching
those of the Gaia data, the first Gaia Data Release (DR1) is 
lacking proper motion data for all stars fainter than the Tycho-2 limit 
of about 11th magnitude.
The upcoming DR2 release, currently scheduled for early 2018 \citep{hsoy} 
will change that.  New proper motions for about 50 times more stars than
contained in TGAS can now be obtained by combining UCAC data at a
mean epoch of 2001 with Gaia position data at epoch 2015.
The accuracy of those proper motions is comparable to those of TGAS
up to about R = 15.0 and provide valuable additional observations
of TGAS stars themselves to noticably improve their proper motions.



\section{UCAC re-reduction}

\subsection{Astrometric solution}

A summary of the relevant features of the UCAC program and data is 
provided in Table 1.
All applicable individual UCAC exposures obtained from Cerro Tololo
Interamerican Observatory (CTIO) and the Naval Observatory Flagstaff 
Station (NOFS) were matched with the Gaia TGAS data which served as 
reference star catalog.
A new astrometric reduction was performed adopting the systematic
corrections of the UCAC $x,y$ data as function of magnitude which were
previously established using the 2MASS \citep{2mass} data covering the
full UCAC magnitude range.  For more details please see the UCAC4
paper \citep{U4}.

However, the systematic errors as a function of location in the
focal plane (the field distortion pattern) as well as the systematic
errors as function of the sub-pixel phase were re-evaluated and
updated using the TGAS residuals separately for the CTIO and NOFS
UCAC data set.
The distortion pattern is mainly determined by the lens and dewar
window but also depends on the actual tilt of the focal plane
with respect to the optical axis which will change after disassemble
and deployment at a new site.
The pixel-phase errors strongly depend on the average width of the
image profiles and thus also change from site to site.

After applying the new corrections, mean residuals as a function of
location in the field of view and sub-pixel phase were reduced to
about 2 mas.
For illustration the new field distortion pattern from the CTIO data 
is shown in Fig.~1.

As before a 6-parameter, linear plate model was adopted for the 
astrometric solution, split into orthogonal ($a$ to $d$) and 
non-orthogonal parameters ($e$ and $f$):
 
 \[  \xi = a x \ + \ b y \ + \ c \ + \ e x \ + \ f y  \]

 \[ \eta = -b x \ + \ a y \ + \ d \ + \ f x \ - \ e y  \]
 
Here $\xi, \eta$ are the standard coordinates (scaled from radian
to arcsec) and $x,y$ the observed center coordinates of star images
on the CCD (scaled from pixel unit to arcsec).

Fig.~2 shows the distribution of the number of reference stars used per
individual UCAC exposure.  The mean is about 50, corresponding to
the average sky density of the TGAS catalog and the UCAC 1.0 square
degree field of view.
The astrometric solution errors are presented in Fig.~3 separately
for the short and long exposure.  These errors include the $x,y$
center errors of the observed image profiles, the reference star
catalog errors and the error contribution from the turbulent
atmosphere.  The latter scales inversely with the square-root of 
integration time and apparently is a significant contribution for 
the short exposures.
The same is shown for UCAC4 data in Fig.~4.
The vast improvement using (almost) error free TGAS reference
stars is striking, showing the high precision of the UCAC observations,
which previously were overshadowed by the Tycho-2 reference star errors.

Fig.~5 shows residuals as a function of calibrated, UCAC bandpass 
magnitude for CTIO data.  This is likely the largest remaining 
contribution to systematic errors in the UCAC data, caused by the poor
charge-transfer efficiency (CTE) of that particular detector \citep{U2}.
No attempt to improve the model was made here because the TGAS
reference stars have a limiting magnitude of about 11.5 while
the UCAC data reaches beyond 16th magnitude.

\subsection{Positions}

Using the above described model, positions of all observed
objects were obtained at their epoch of observation for all
applicable CCD exposures. 
Fig.~6 shows the mean observed epoch as a function of declination.
The UCAC survey began in the south and ended at the north celestial pole.

Mean observed UCAC5 positions were obtained from a weighted
mean of the individual positions (images).  Outliers were rejected
as much as possible with the small number of observations available
for this task.
Fig.~7 shows the distribution of the number of observations used
for the mean positions.  
In cases with a total of 2 images which display discrepant positions, 
the outlier can not be identified and an unweighted mean position 
is given with the number of used images set to zero.

Due to the small field of view and the desire to cover all-sky, the
number of observations per star is small. A 2-fold (center-in-corner)
overlap pattern was adopted, with a short and long exposure
on each field to extend the dynamic range.  
Thus stars in the 8 to 9.5 mag range typically have
2 images (from the short exposures only), stars in the 9.5 to
about 14.5 mag range usually have 4 images (short and long
exposure), while stars fainter than about R = 14.5 show up
only on the long exposures.

The errors on the UCAC5 positions were obtained from the formal
errors ($x,y$ center errors plus astrometric solution error 
propagation) of the individual images used for the mean position. 
A 10 mas term was added in quadrature to account for 
possible remaining systematic errors and to provide a more realistic
error floor for the small number statistics of individual stars.
This error floor is likely dominated by the remaining systematic
errors as a function of magnitude and the same value as in the
earlier UCAC4 reductions was adopted here because no changes in
the calibration model for those errors were made.

\subsection{Adding NOMAD data to Gaia DR1}

The Naval Observatory Merged Astrometric Dataset (NOMAD) catalog
\citep{nomad} contains about a billion entries, covers all-sky, and
the magnitude range from naked eye stars to about V = 20 mag.
It contains positions, proper motions, optical and near IR magnitudes.
NOMAD positions were updated to the 2015 epoch using its proper
motions and then matched to the Gaia DR1 data.
Over 638 million sources were matched within 1 arcsec based on
position only.  A new, internal catalog was created, which 
adds NOMAD data, if available, to all sources in Gaia DR1.

These new proper motions are of value for stars not contained
in the UCAC data, thus mainly for stars fainter than about R = 16 mag.
The new proper motion data for stars in common with UCAC data
are very helpful to correctly match UCAC stars to the Gaia data,
particularly for stars with moderate to large proper motions
(see below).

\section{Results}

\subsection{Position error analysis}

Having results from largely different exposure times allows us to
determine the individual error contributions of the UCAC observations.
The astrometric solution error, $\sigma_{S}$, has 3 components,
the errors of the reference star catalog $\sigma_{r}$ at
epoch of observations, the $x,y$ data error $\sigma_{xy}$,
and the error introduced by the turbulence in the atmosphere,
$\sigma_{a}$, 

\[ \sigma_{S}^{2} \ = \ \sigma_{r}^{2} \ + \ \sigma_{xy}^{2} \
 + \sigma_{a}^{2}  \]

The variance of the error contribution from the atmosphere scales
inversely with exposure time, $t$,

\[ \sigma_{a}^{2} \ = \ \sigma_{0}^{2} / t \]

while the other 2 error contributions are independent
of exposure time.  The reference star errors in the magnitude
range used here (mostly G = 9 to 11) are mainly a function of
epoch (see below), and the $x,y$ errors are nearly constant
for these high S/N data.  We define

\[ \sigma_{c}^{2} \ = \ \sigma_{r}^{2} \ + \ \sigma_{xy}^{2} \]

leading to

\[ \sigma_{S}^{2} \ = \ \sigma_{c}^{2} + \sigma_{0}^{2} / t \]

For 2 different exposure times we thus have 2 linear equations
with known $\sigma_{S}$ to directly solve for $\sigma_{c}$, and 
$\sigma_{0}$.
From Fig.~3 we see the peaks of the UCAC5 astrometric solution errors,
$\sigma_{S}$ at 29 mas for the short exposures (on average 25 sec), 
and 19 mas for the long exposures (on average 125 sec).  The peak values
of these distributions can be interpreted as ``typical good quality"
observations.
With these numbers we get $\sigma_{0}$ = 122 mas sec$^{1/2}$,
the error contribution from the atmosphere, 
and $\sigma_{c}$ = 15.5 mas, the RMS combined error of the reference
stars and the $x,y$ centroiding error.

The reference star errors at the mean UCAC epoch of 2001 are
dominated by proper motion errors.  From \citep{G4} we see that
for TGAS stars not in the Hipparcos sample the median proper motion errors
are about 1.1 and 0.9 mas/yr for the RA and Dec component, respectively.
Assuming 1.0 mas/yr here and an average epoch difference of 14 years 
between the UCAC and the TGAS positions, we have an estimate
of $\sigma_{r}$ = 14 mas.
This allows us to solve for the last remaining error contribution,
$\sigma_{xy}$ = 6.7 mas.

This error includes random centroiding errors as well as remaining,
uncorrected systematic errors for example as a function of pixel phase,
location in the field of view, and magnitude (over the range of the
reference star magnitudes).
Note, $\sigma_{xy}$ is the average position error of high 
S/N UCAC observations per coordinate and per exposure (which is 1/135 
pixel).
Assuming the error contributions of the reference stars and those 
of the atmosphere are small, the obtained observational precision
then is limited by $\sigma_{xy}$.  The former will soon be obtained 
with the 2nd Gaia data release, and the latter can be achieved with 
long integration times. 

The image center algorithm used to derive the UCAC $x,y$ data from
the pixel data is a weighted least-square fit with a 2-dimensional
Gaussian model followed by extensive empirical modelling of the
pixel-phase and other systematic errors.  
No elaborate PSF fitting was performed, instead the high precision
was obtained by analyzing and modeling the position residuals.
Details are described in \citep{U2}.

\subsection{UCAC5--Gaia proper motions}

The Gaia plus NOMAD data catalog positions were propagated to a mean 
epoch of 2001 and then matched with the UCAC5 observational catalog.
Again, a match was assumed if the position difference is within 
1 arcsec in each coordinate.  This resulted in 107.7 million stars
in common between the UCAC and Gaia data.  New proper motions were
calculated for all those stars based on the 2 epoch positions (Gaia
at 2015, UCAC5 at about 2001) with a mean epoch difference of about 
14 years.  No NOMAD positional data were used for these proper motions.
The NOMAD data only served to facilitate the proper match, thus
providing also results for stars with large proper motion.

Errors of the proper motions were obtained from the formal position
errors of Gaia and UCAC5 and the epoch difference of individual stars.
Fig.~8 shows the distribution of the proper motion errors.
This distribution peaks at about 1.2 mas/yr which is comparable to 
the TGAS proper motion errors, but for many millions of more stars.

Fig.~9 shows the strong dependence of our proper motion errors
with brightness.  Errors in proper motion slowly increase to
2.5 mas/yr at about magnitude 15 and then rapidly increase to
about 10 mas/yr at the limiting magnitude of 16.5 due to the
low S/N of faint stars.

Table 2 lists the data items of our UCAC5 catalog.
Positions are given at the mean UCAC observed epoch for each
star on the Gaia reference frame. 
The UCAC5 binary data file, sorted by declination is 4.3 GB large 
and will be available from the Centre de Donn{\'e}es
astronomiques de Strasbourg (CDS).

\subsection{Close doubles}

A match of UCAC5 with itself was performed revealing some 52,000
multiple matches within 2 arcsec.  Spot checks indicate 2 types
of cases. The first case are close pairs in Gaia DR1, i.e. 2 real 
stars which are matched to the same UCAC observation i.e. the
photocenter of the pair which is unresolved in the UCAC data.
The second case are close doubles seen as 2 stars in both, Gaia and 
UCAC data.  No stars identified in this investigation were removed
from the published catalog.
The Gaia DR1 source identifier remains unique within the UCAC5 
catalog, because the DR1 entries were used as initial input list.
However, in a few cases the same UCAC object is matched to 
2 different DR1 entries.
No duplicate entries, i.e. entries with positions identical
to within a few mas were found in UCAC5.

\subsection{Comparison to TGAS}

A separate, formatted data file is available for the 2,054,491 stars
in common between TGAS and UCAC5, which also lists the differences
in proper motion (UCAC5 - TGAS).
The TGAS proper motions use the Gaia observations (at epoch 2015)
together with the Hipparcos observations of these stars (at epoch 1991).
The UCAC5 proper motions are based on the astrograph observations at
a mean epoch of about 2001 and the Gaia 2015 observations.
Of course these proper motions are somewhat correlated,
both use the same later epoch data but different early epoch
observations.  Furthermore, the UCAC5 data uses TGAS stars as
reference stars which include use of TGAS proper motions.
However, the UCAC5 epoch observations are largely independent,
new observations due to the fact that typically 20 to 200 such
reference stars are used in the astrometric solutions of UCAC
observations with a simple, linear model of 6 parameters,
which provide a large degree of overdetermination in the 
least-squares reductions.

The distribution of UCAC5 proper motions is shown in Fig.~10.
The distribution of the differences in proper motions are small (Fig.~11).
The formal errors of proper motions for stars in common are
shown in Fig.~12 and 13 for the TGAS and UCAC5 data, respectively.
Both proper motions are similar in performance.
TGAS has more stars with about 1 mas/yr or less errors, while
the UCAC5 proper motions are somewhat better for
stars with about 2 mas/yr TGAS proper motion errors and above.

\section{External comparisions}

\subsection{Star Cluster}

As an example of astrophysical application and validation of the
UCAC5 proper motions, 2 cluster areas were picked with a box size
of 30 arcmin.  In Fig.~14 UCAC5 proper motions of all such stars in 
the open cluster NGC 3532 area are shown in comparison to the 
fourth release of the Southern Proper Motion (SPM4) program
\citep{spm4} proper motions, the previous ``gold standard"
for absolute proper motions of faint stars.
Fig.~15 shows the same for the area around the globular cluster NGC 6397.
A significant improvement in the ability to separate cluster member stars 
from non-cluster stars is seen with the UCAC5 proper motions as
compared to the SPM4 proper motions.

\subsection{Extragalactic Sources}

Due to their extreme distance, extragalactic sources will have negligible
proper motions. Thus the observed proper motions show the limitations
of the catalog data.
A match of UCAC5 with LQAC3 \citep{lq3} was performed,
which lists over 321,000 confirmed extragalactic sources, mostly QSOs.
A total of 2001 LQAC3 sources are in common with the UCAC5 within
1.5 arcsec.  

Table 3 summarizes results for selected sub-sets, per coordinate and
for the proper motions in mas/yr as well as for their normalized
values (proper motion divided by formal error of the proper motion
of that object).
Subsets were selected by minimum number of UCAC observations used
for the mean UCAC5 position (nu), the LQAC3 catalog object type
(R = astrometric radio source, Q = QSO, others include AGN and
BL-Lac objects), the UCAC5 bandpass mean observed magnitude (mag),
and the redshift $z$.  The total number of objects for each set
is also given, while 80\% of these are used to derive the mean and
RMS results, excluding the top and bottom 10\% of the data after 
sorting (to prevent outliers to affect this analysis).
The entry ``all" in Table 3 means no restriction has been applied
for that particular column item.
The last 2 lines in Table 3 list results for sources in common with
the 2nd version of the International Celestial Reference Frame (ICRF2).
The ICRF2 \citep{icrf2} currently defines the inertial coordinate system
on the sky, and is derived from highly accurate, radio, very long 
baseline interferometry (VLBI), observations of compact, extragalactic 
sources.

For most data sets a small, negative offset in the mean UCAC5 proper
motions of about $-$0.5 mas/yr is seen, while the ICRF sources do
not show this.
The RMS scatter of the observed proper motions is larger than the
formal errors for most data sets.
However, when excluding low redshift sources the RMS scatter is
significantly reduced.


\section{Discussion and Conclusions}

High quality, ground-based observations to obtain positions up to
now were impacted by relatively poor reference star data. The TGAS
catalog now shows the full potential of such observations. 
Future Gaia data with even more accurate positions and for many 
more, fainter stars will have a big impact on such ground-based
observations for example in time-domain astronomy.
The quality of the telescope and instrument will be more
important than ever before.
The limitation of the USNO redlens astrograph data can be expected
to be significantly below 1/100 of a pixel per coordinate for single,
long exposure observations of high S/N stars.


The UCAC5 positions on the Gaia coordinate system provide additional
data of similar quality to the Hipparcos mission Tycho star observations
and thus have the potential to improve the TGAS proper motions.
UCAC5 provides new, accurate proper motions for millions of more stars 
fainter than TGAS, which will allow astronomers to have a preview
into research possible only with the next Gaia data release.
At the faint end UCAC5 proper motion errors are relatively large
due to the low S/N ratio of these observations.
Better proper motions for stars fainter than about 15th mag are
available from proper motions obtained by combining NOMAD with Gaia DR1 
(catalog of 503 million stars is available upon request), or the
recently published PPMXL re-reduction, called the HSOY \citep{hsoy}
catalog.

The biggest issue with the UCAC5 data remains the problematic
corrections of systematic errors as a function of magnitude due to
the poor CTE of the CCD used in the program.
However, the remaining systematic positional errors are expected
to not exceed 10 mas over the entire magnitude range
of UCAC data (verified with TGAS data for the 8 to 11 mag range),
which can lead to systematic errors in the UCAC5 proper motions
up to 0.7 mas/yr.  This is confirmed with comparisons to
extragalactic sources.

Restricting the sample of extragalactic sources to redshift of
0.5 or higher results in a significant reduction in the UCAC5
observed proper motion scatter.  
This is another indication for possible optical structure of
nearby extragalactic source affecting the observed image centers.
Both epochs (UCAC and Gaia) used for these proper motions are
based on optical data, however, the resolution of Gaia is at
least 10 times higher than that of the astrograph.
It appears that both instruments ``see" a different photocenter
at least for cosmological nearby sources, resulting in a larger
scatter of the proper motions, despite the fact that they are
typically brighter with smaller formal position errors
than more distant sources.



\acknowledgments

We used the PostgreSQL Q3C sky-indexing scheme for some of our external
catalog comparisons \citep{q3c}, as well as the Department of Defense
Celestial Database of the USNO Astrometry Department, developed by 
V.Makarov, C.Berghea, and J.Frouard.
Pgplot by California Institute of Technology was used to produce plots.
The gfortran and g77 compilers were used for code development.

\clearpage


\begin{figure}
\includegraphics[angle=-90,scale=0.45]{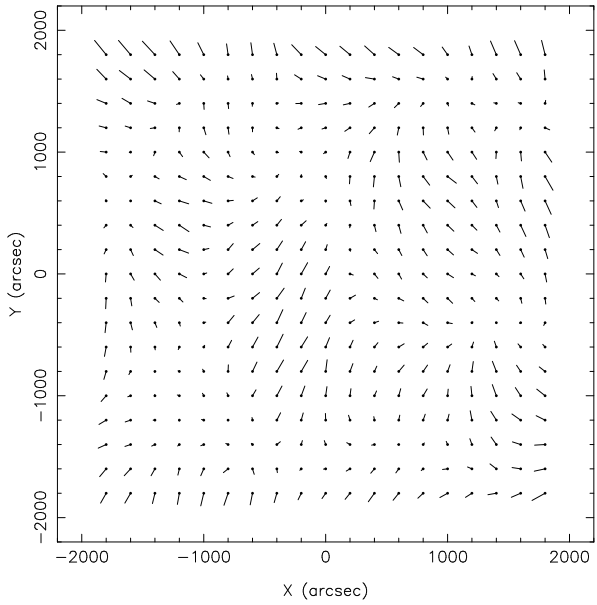}
\caption{Field distortion pattern of the UCAC instrument from
TGAS reference star residuals of all applicable exposures taken
at CTIO. The vectors are scaled by a factor of 5000. The largest
residual vectors are about 25 mas long.}
\end{figure}

\begin{figure}
\includegraphics[angle=0,scale=0.45]{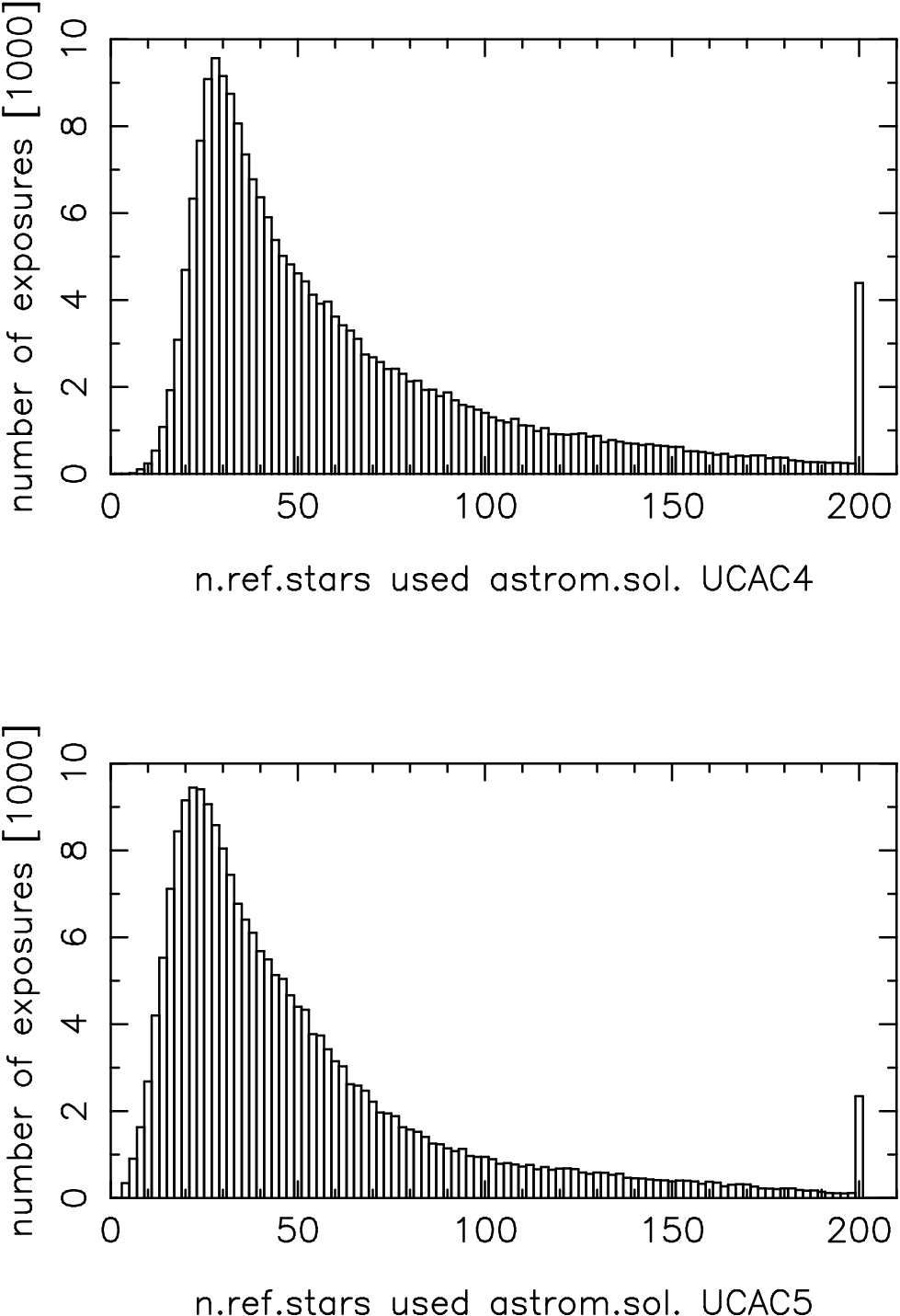}
\caption{Distribution of the number of reference stars used in the
UCAC5 astrometric reductions (TGAS stars) per individual exposure.
The last bin sums up all exposures with 200 and more reference stars.}
\end{figure}

\begin{figure}
\includegraphics[angle=0,scale=0.45]{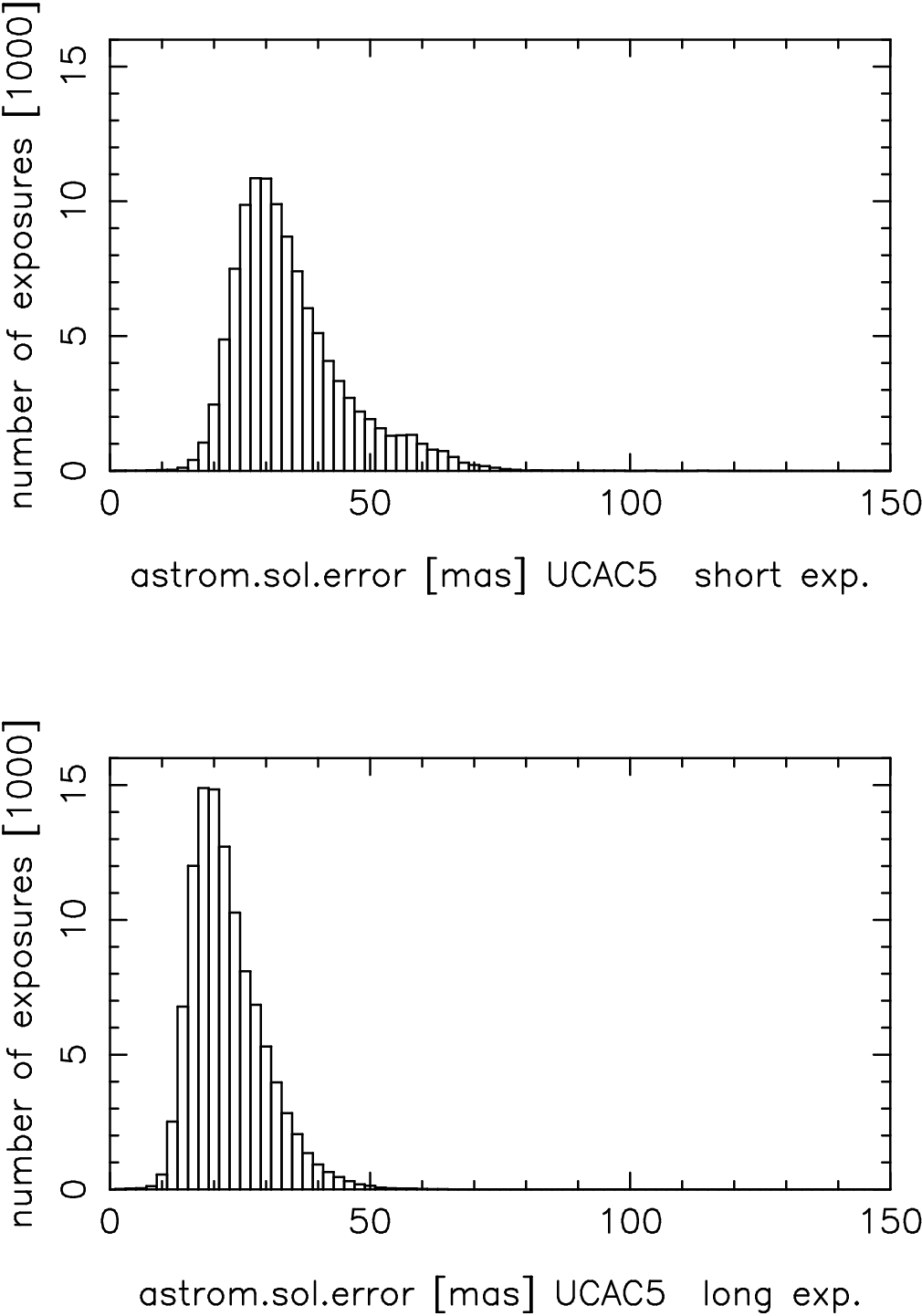}
\caption{Distribution of astrometric solution error of UCAC5
individual exposures using TGAS reference stars and the new
systematic error model.
The top diagram shows the results for the short exposures 
(20, 25, 30 sec) while results for the long exposures (100, 125, 150 sec)
are shown in the histogram at the bottom.}
\end{figure}

\begin{figure}
\includegraphics[angle=0,scale=0.45]{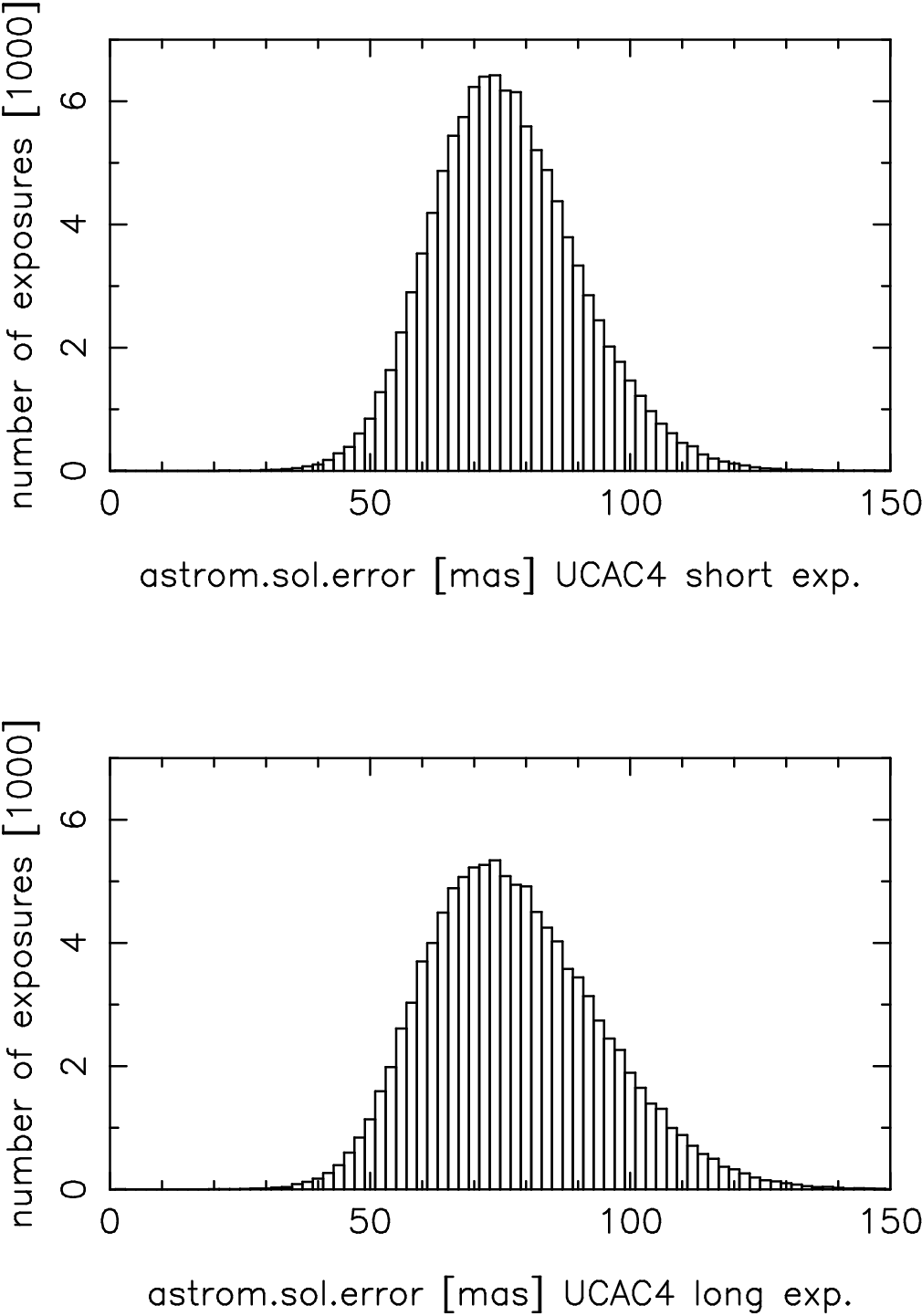}
\caption{Same as the previous figure but for the UCAC4 results
using Tycho-2 reference stars.}
\end{figure}


\begin{figure}
\includegraphics[angle=-90,scale=0.35]{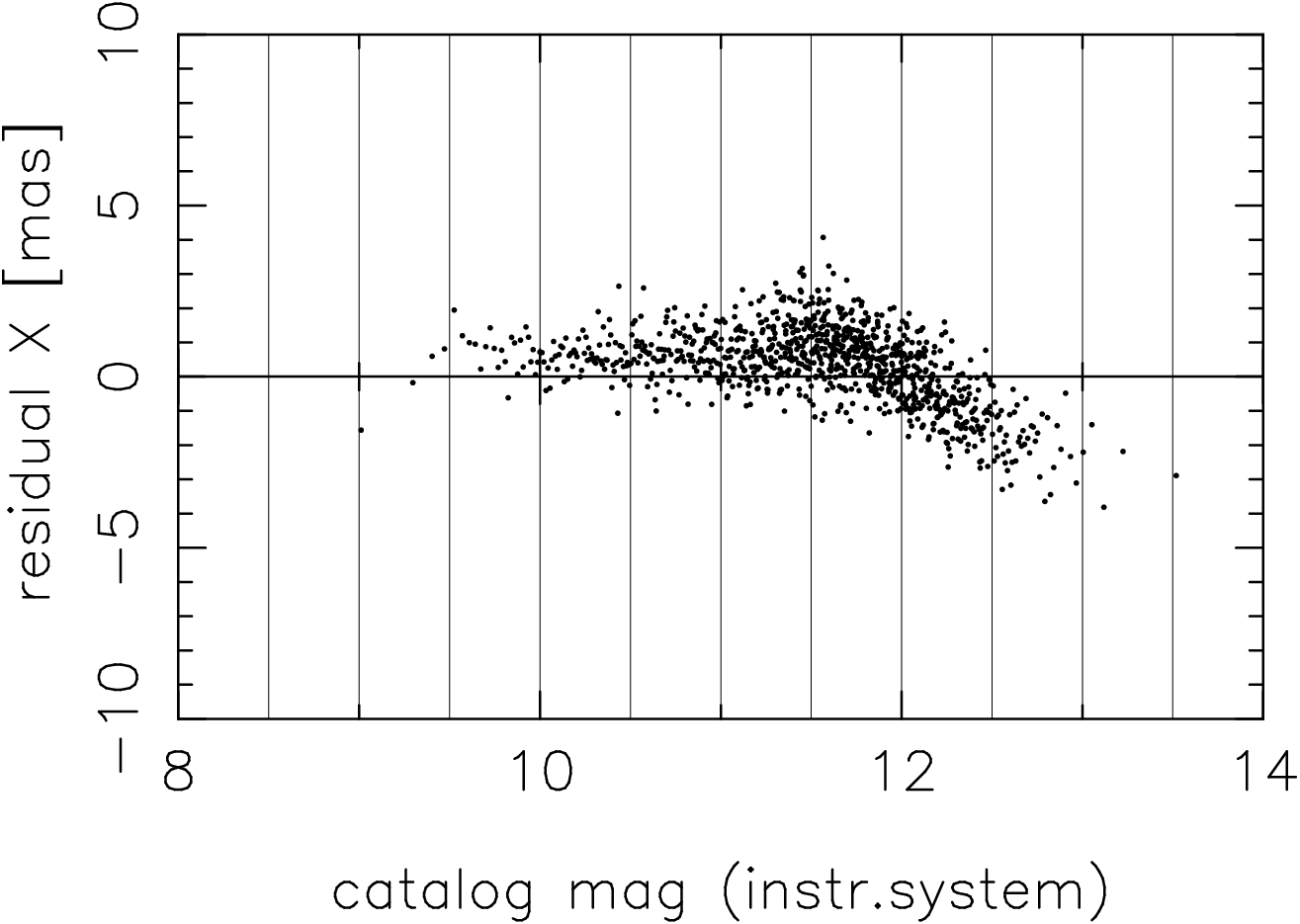} 
\includegraphics[angle=-90,scale=0.35]{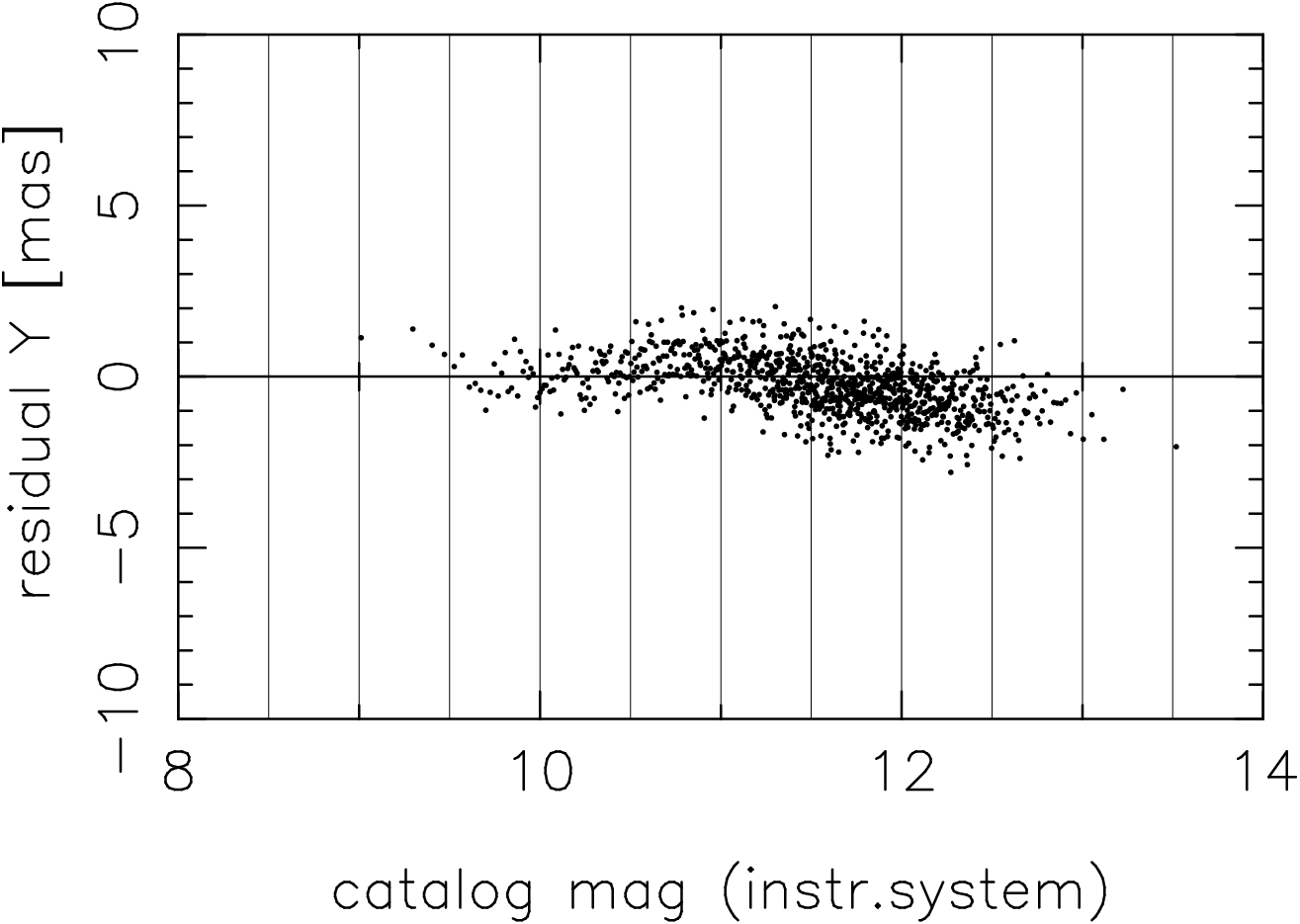} 
\caption{UCAC5 residuals (along $x$ = RA top, along $y$ = Dec
bottom) as a function of calibrated UCAC bandpass magnitude 
from data taken at CTIO (long and short exposures together). 
Each dot is the mean of 5000 residuals.} 
\end{figure}

\begin{figure}
\includegraphics[angle=-90,scale=0.30]{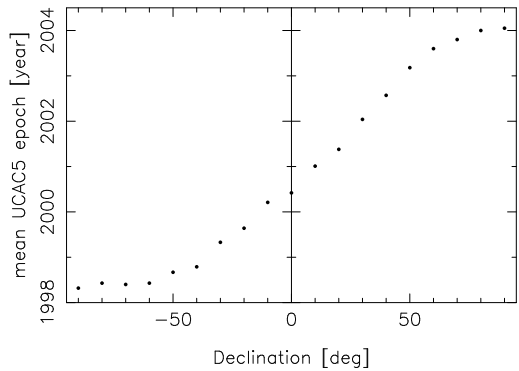}
\caption{UCAC5 mean observing epoch as a function of declination.}
\end{figure}

\begin{figure}
\includegraphics[angle=-90,scale=0.35]{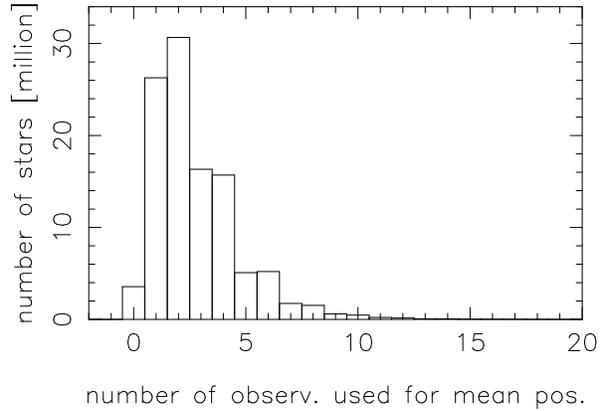}
\caption{Distribution of number of observations (individual
exposures) used for the mean UCAC5 position. For objects
in the zero bin see text.}
\end{figure}

\begin{figure}
\includegraphics[angle=0,scale=0.45]{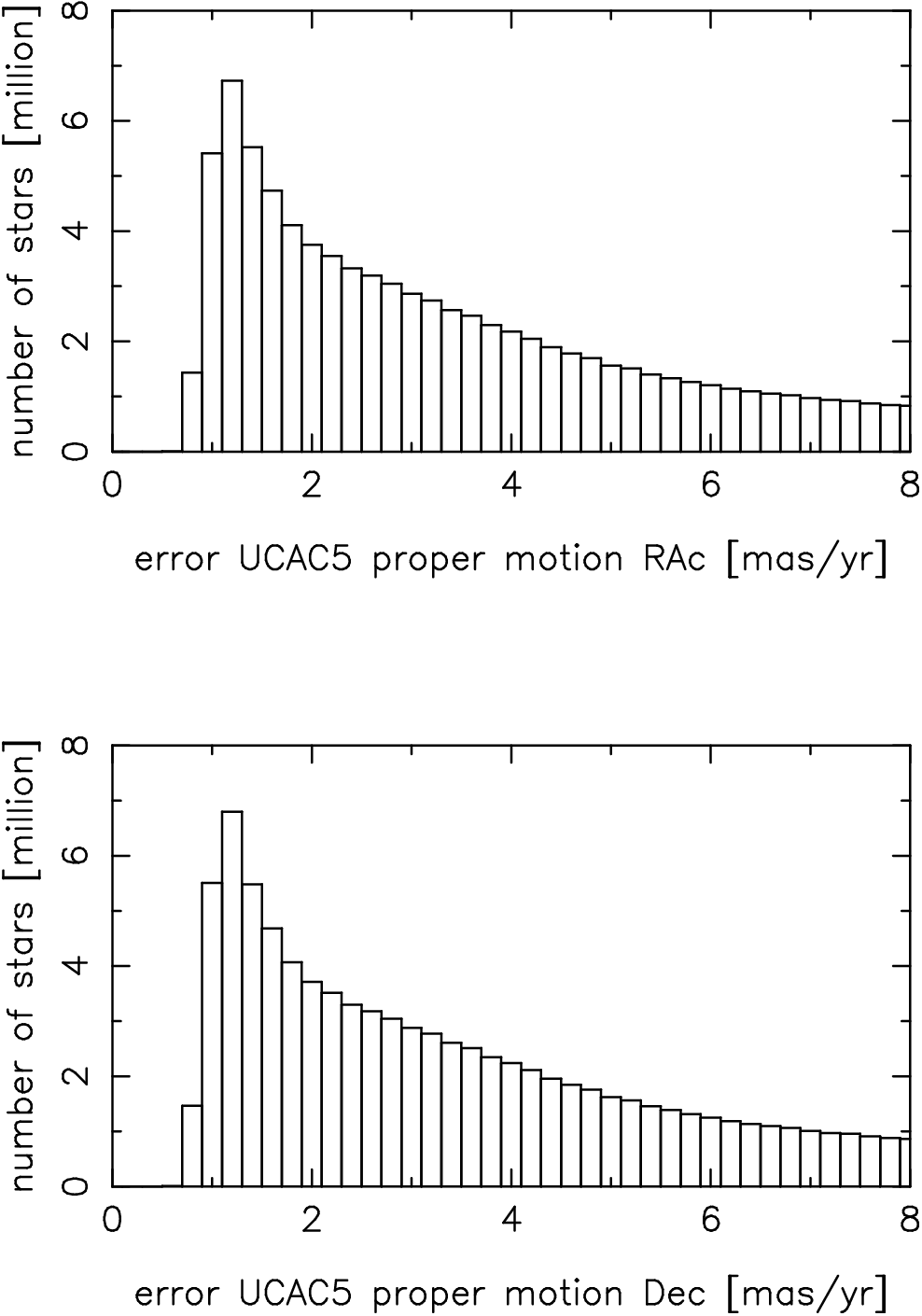}
\caption{Distribution of errors in proper motion of our
UCAC5-Gaia catalog (RA on top, Dec on the bottom).}
\end{figure}

\begin{figure}
\includegraphics[angle=-90,scale=0.35]{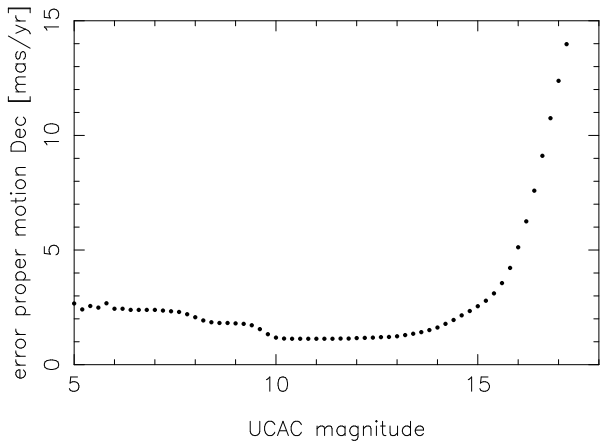}
\caption{Errors of proper motions of the UCAC5-Gaia catalog
as function of UCAC magnitude.}
\end{figure}

\begin{figure}
\includegraphics[angle=-90,scale=0.35]{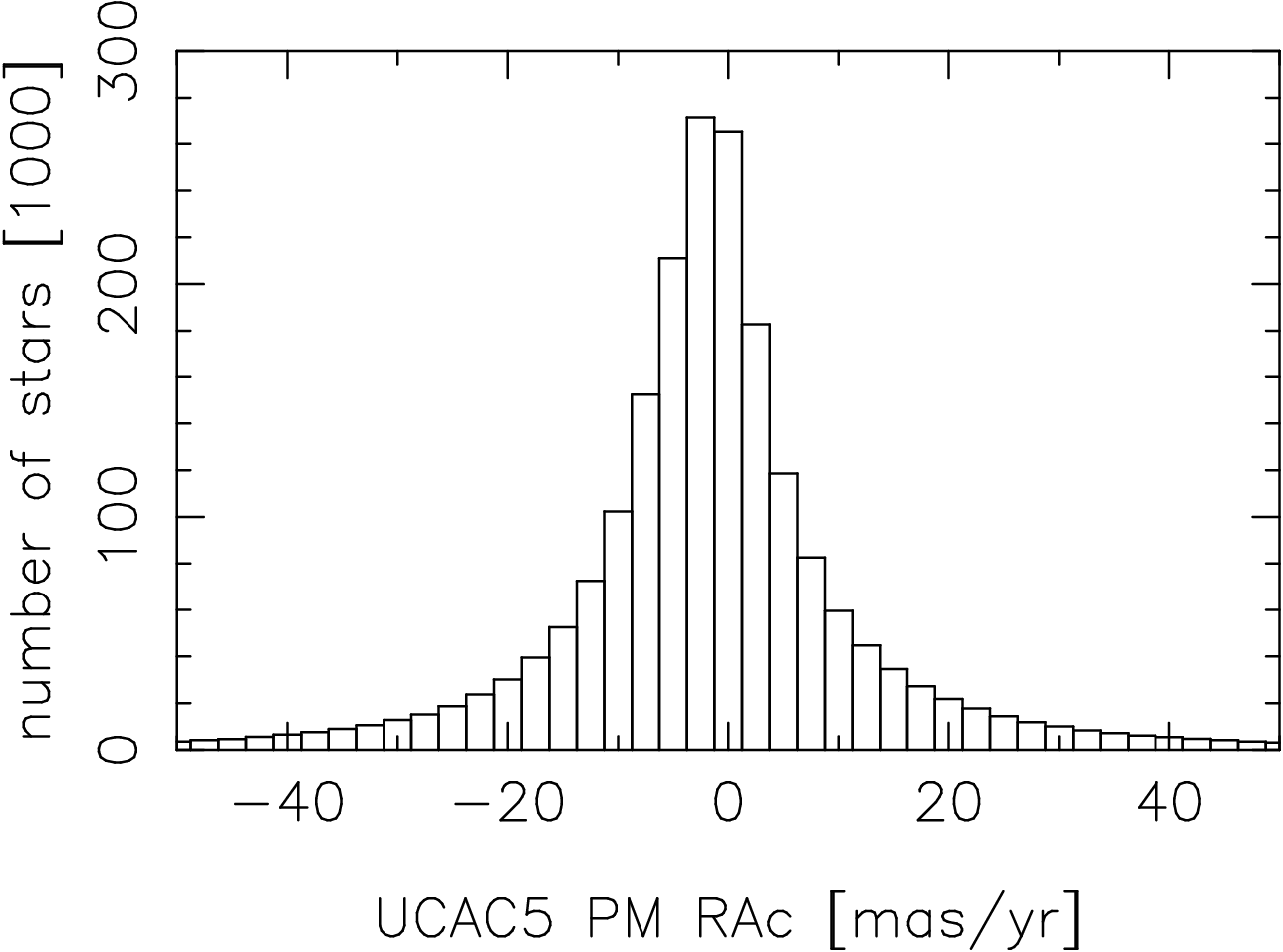}
\includegraphics[angle=-90,scale=0.35]{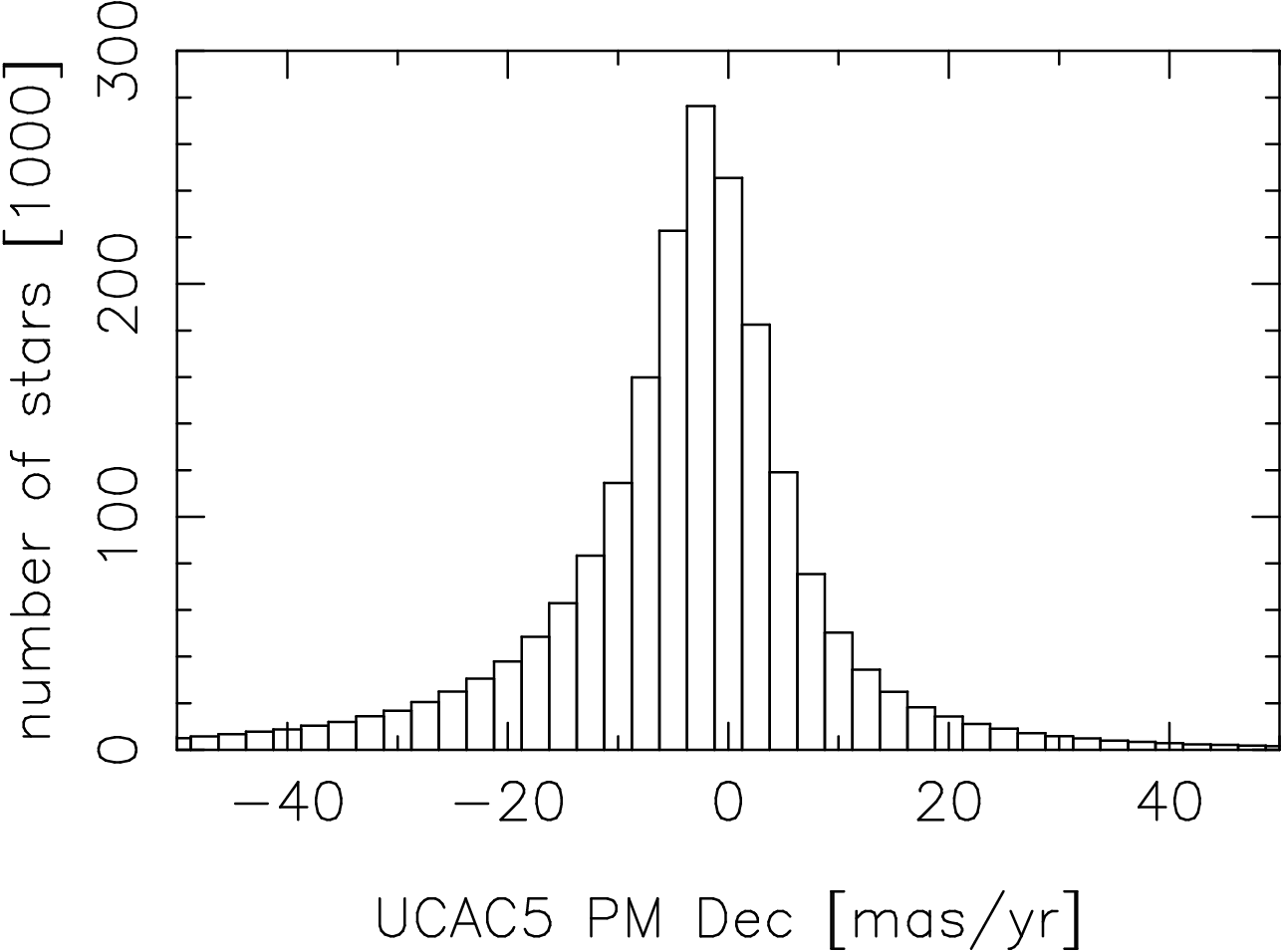}
\caption{Distribution of UCAC5 proper motions of stars in common
between UCAC5 and TGAS, for RA (top), and Declination (bottom).}
\end{figure}

\begin{figure}
\includegraphics[angle=-90,scale=0.35]{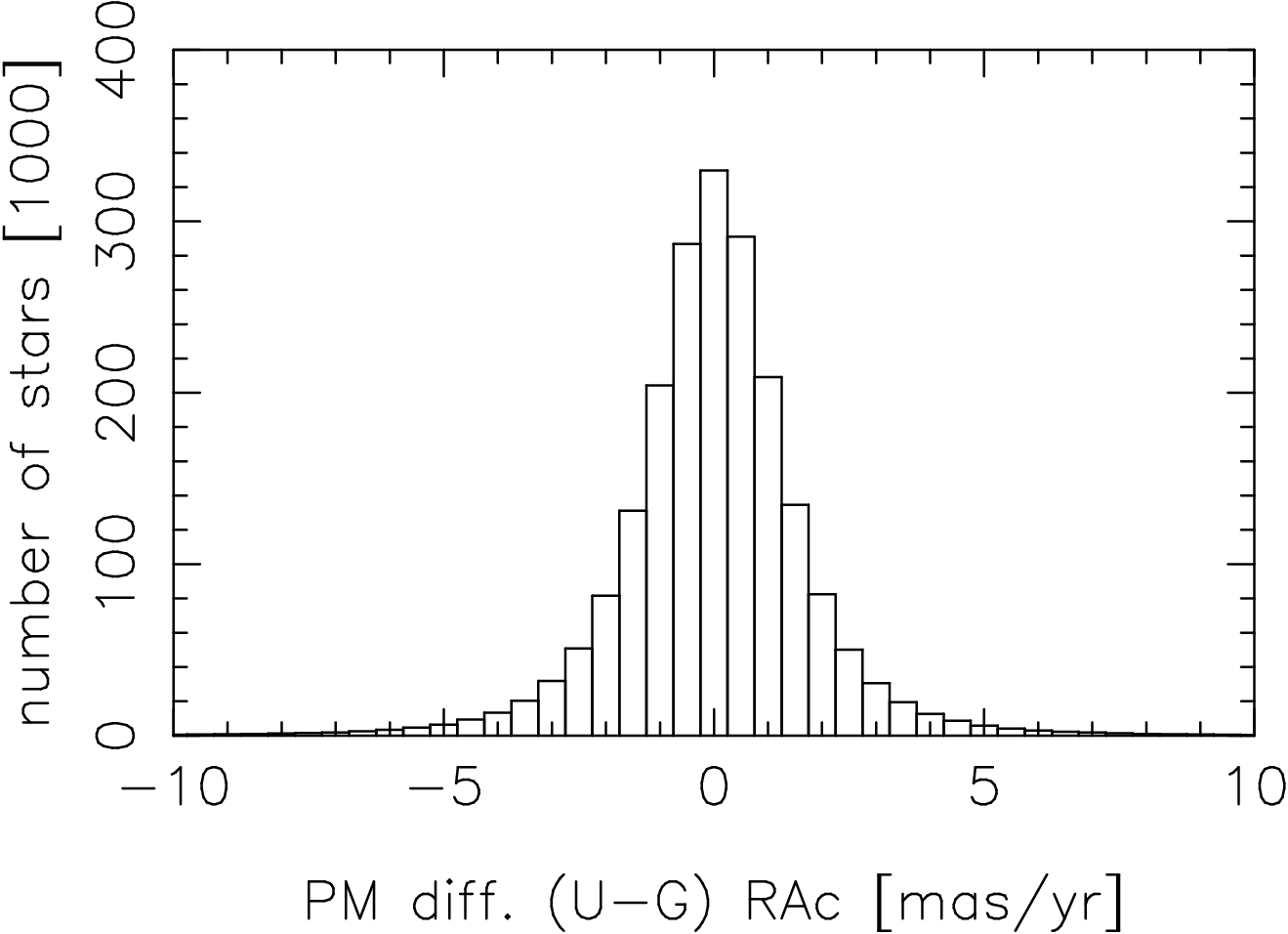}
\includegraphics[angle=-90,scale=0.35]{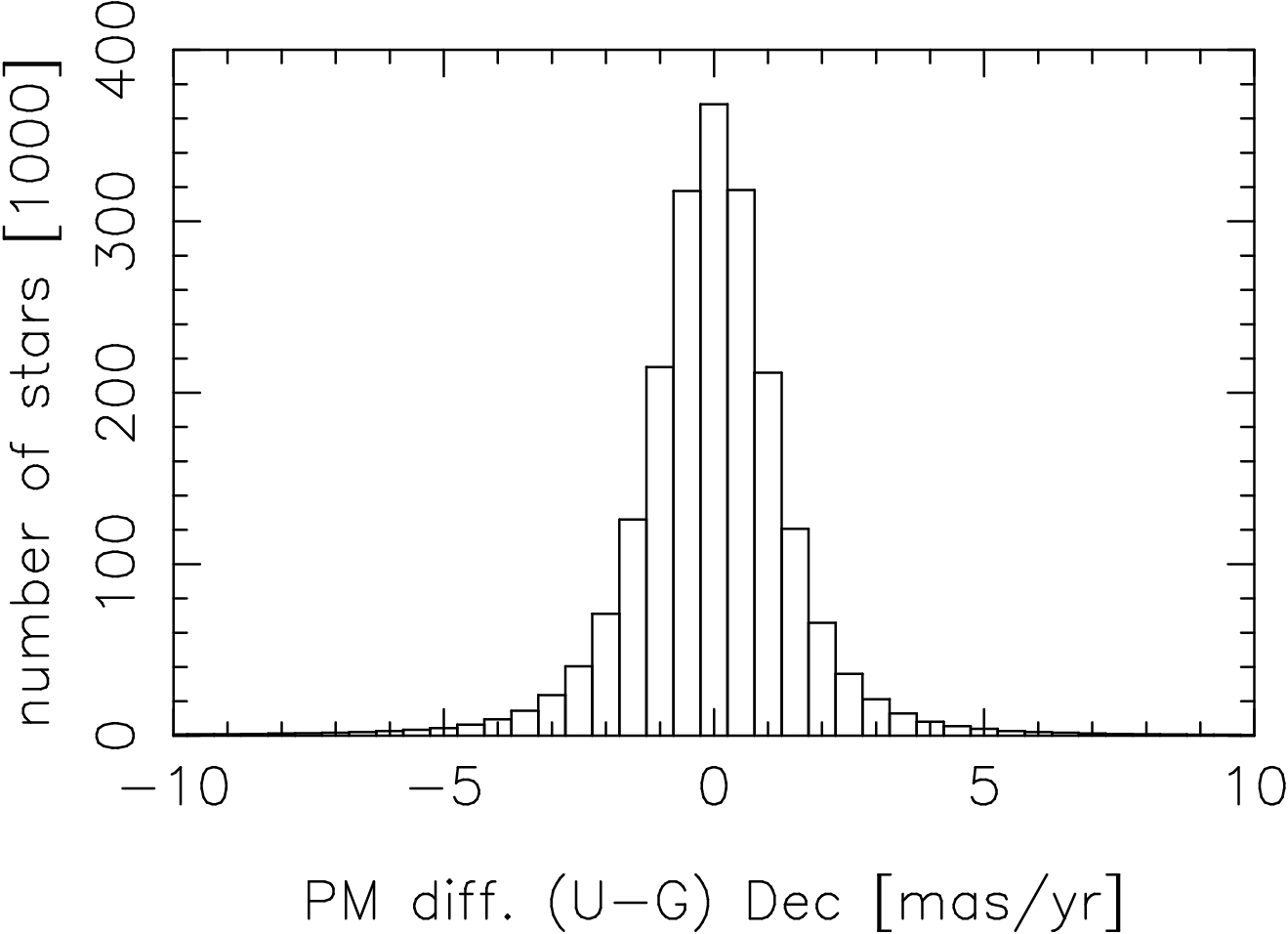}
\caption{Differences between TGAS and UCAC5 proper motions
for RA (top) and Declination (bottom).}
\end{figure}

\begin{figure}
\includegraphics[angle=-90,scale=0.35]{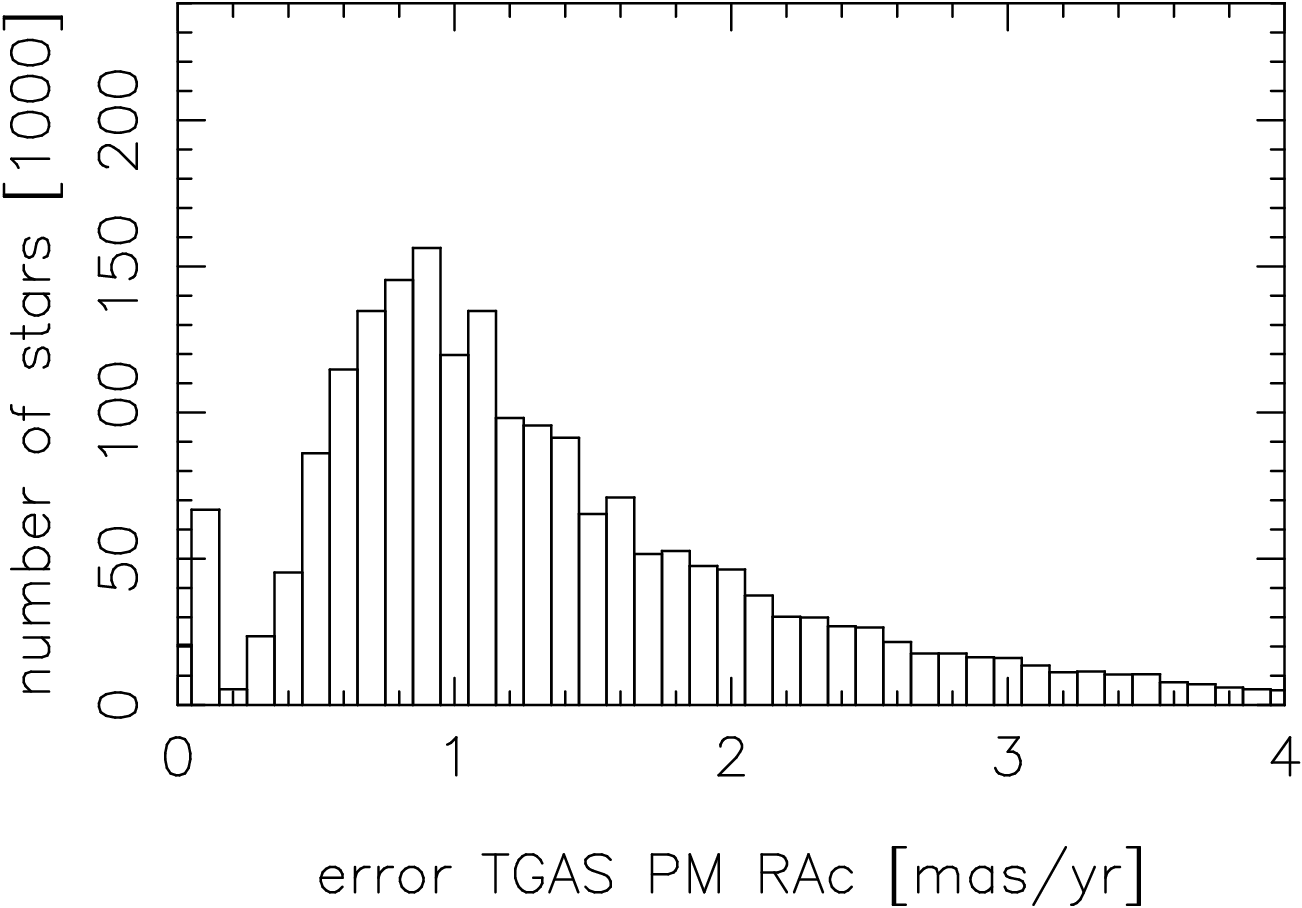}
\includegraphics[angle=-90,scale=0.35]{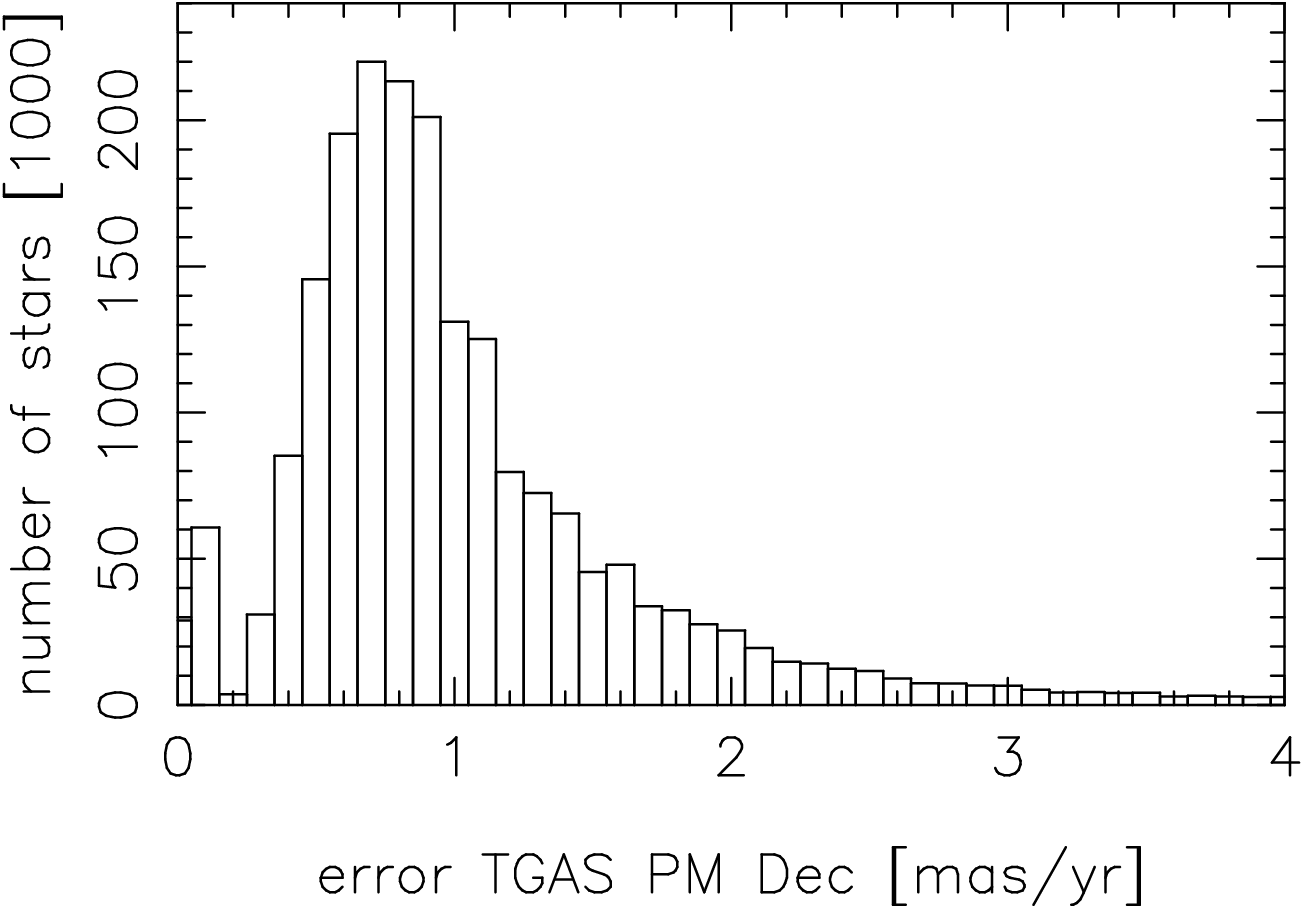}
\caption{Formal errors of TGAS proper motions of stars in common
between UCAC5 and TGAS, for RA (top), and Declination (bottom).}
\end{figure}

\begin{figure}
\includegraphics[angle=-90,scale=0.35]{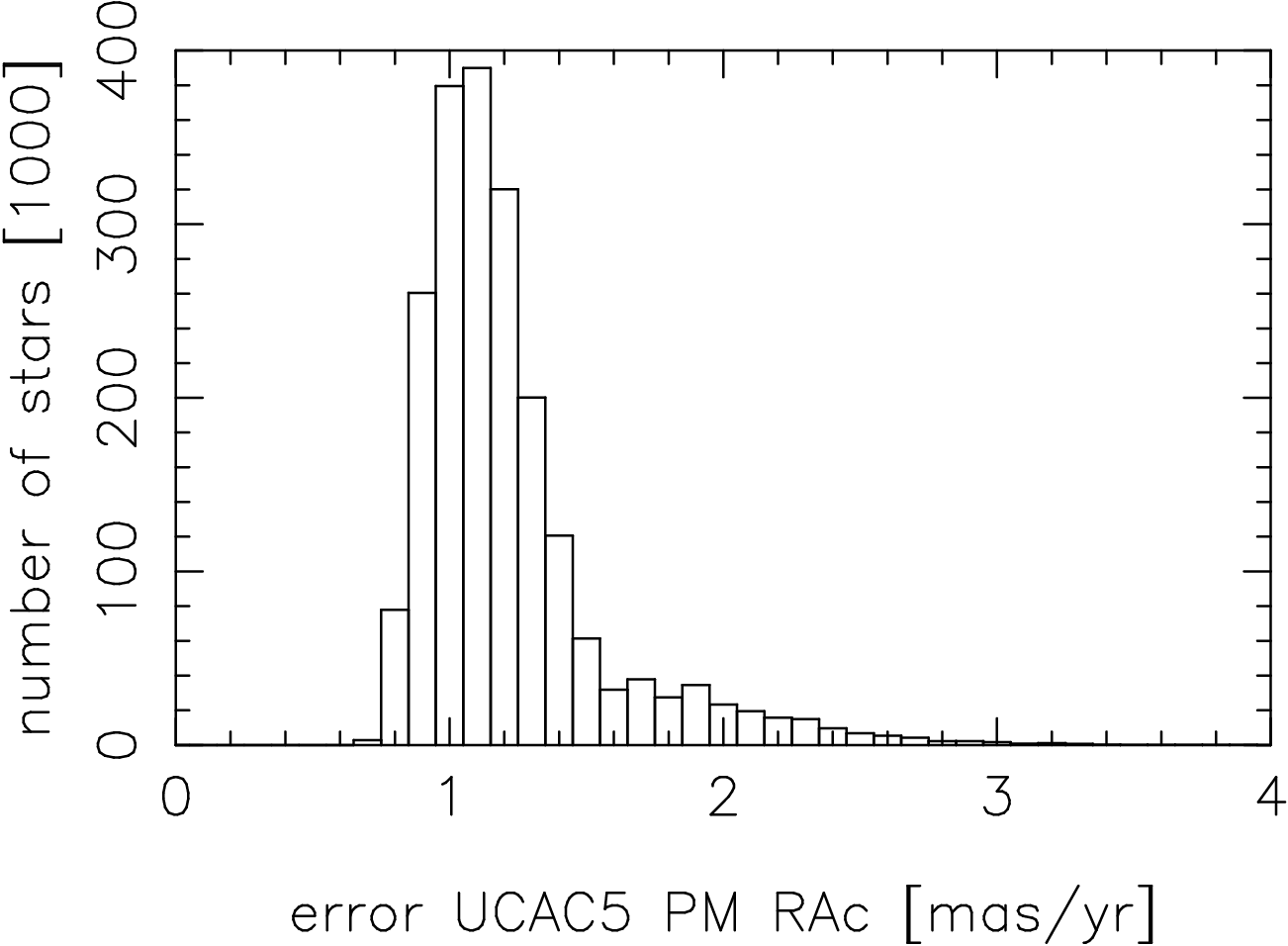}
\includegraphics[angle=-90,scale=0.35]{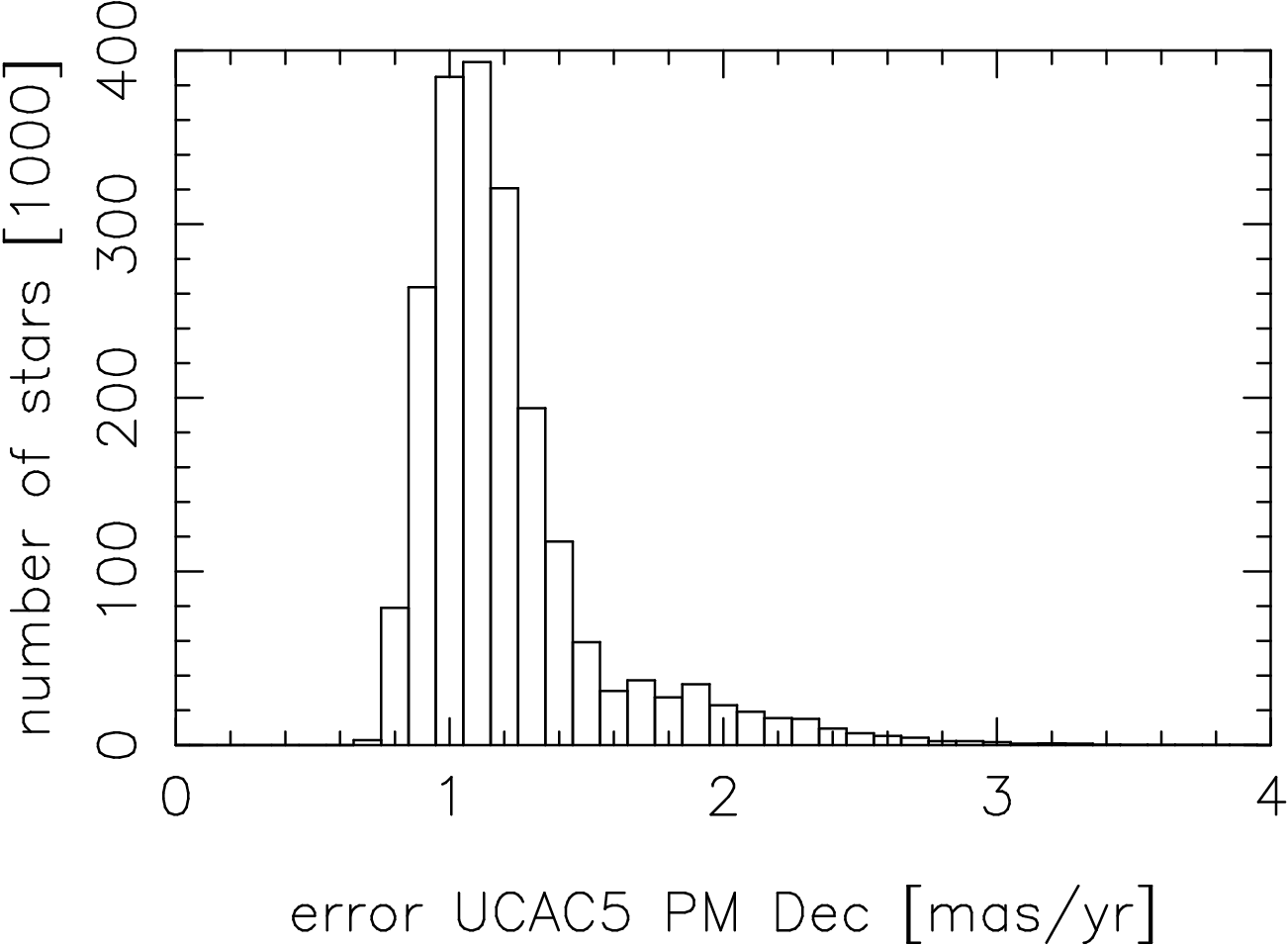}
\caption{Formal errors of UCAC5 proper motions of stars in common
between UCAC5 and TGAS, for RA (top), and Declination (bottom).}
\end{figure}

\begin{figure}
\includegraphics[angle=0,scale=0.70]{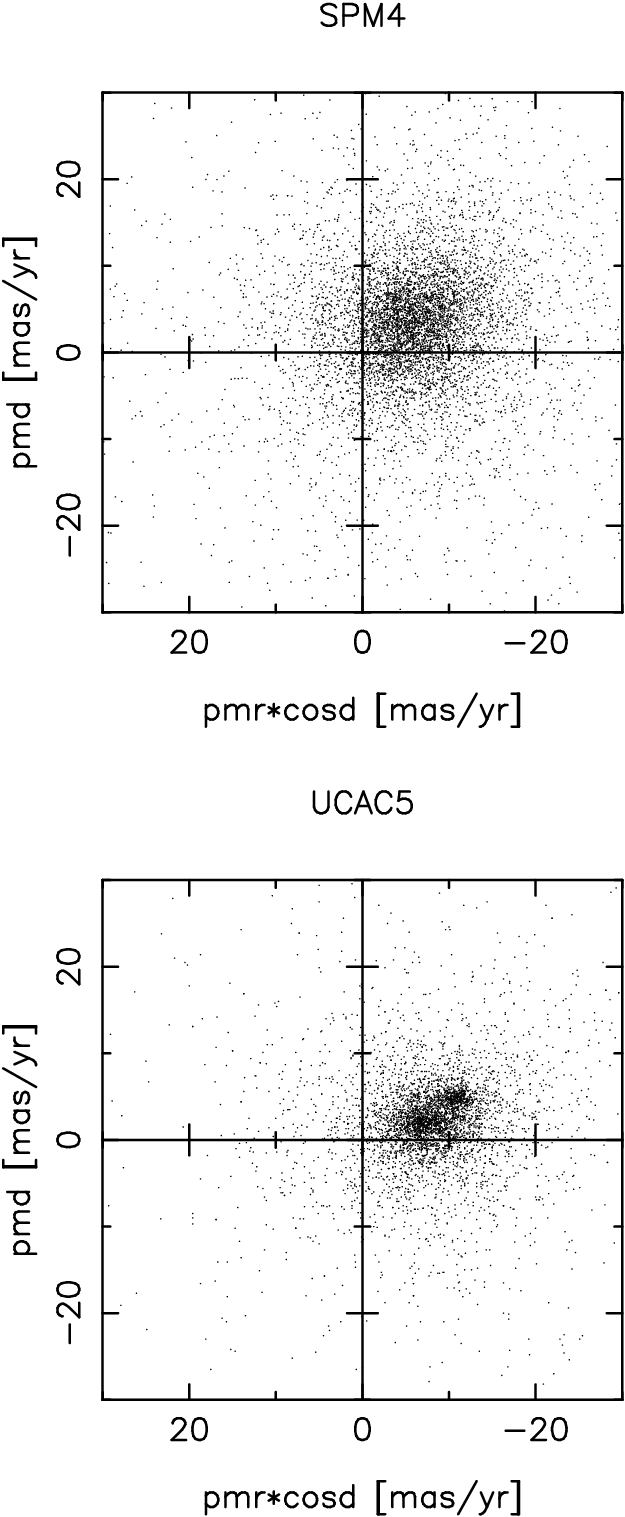}
\caption{SPM4 (top) and UCAC5 (bottom) proper motions 
(pmr = along RA, pmd = along Dec) of stars in common between UCAC5 
and SPM4 in the 30 arcmin area around the open cluster NGC 3532.}
\end{figure}

\begin{figure}
\includegraphics[angle=0,scale=0.70]{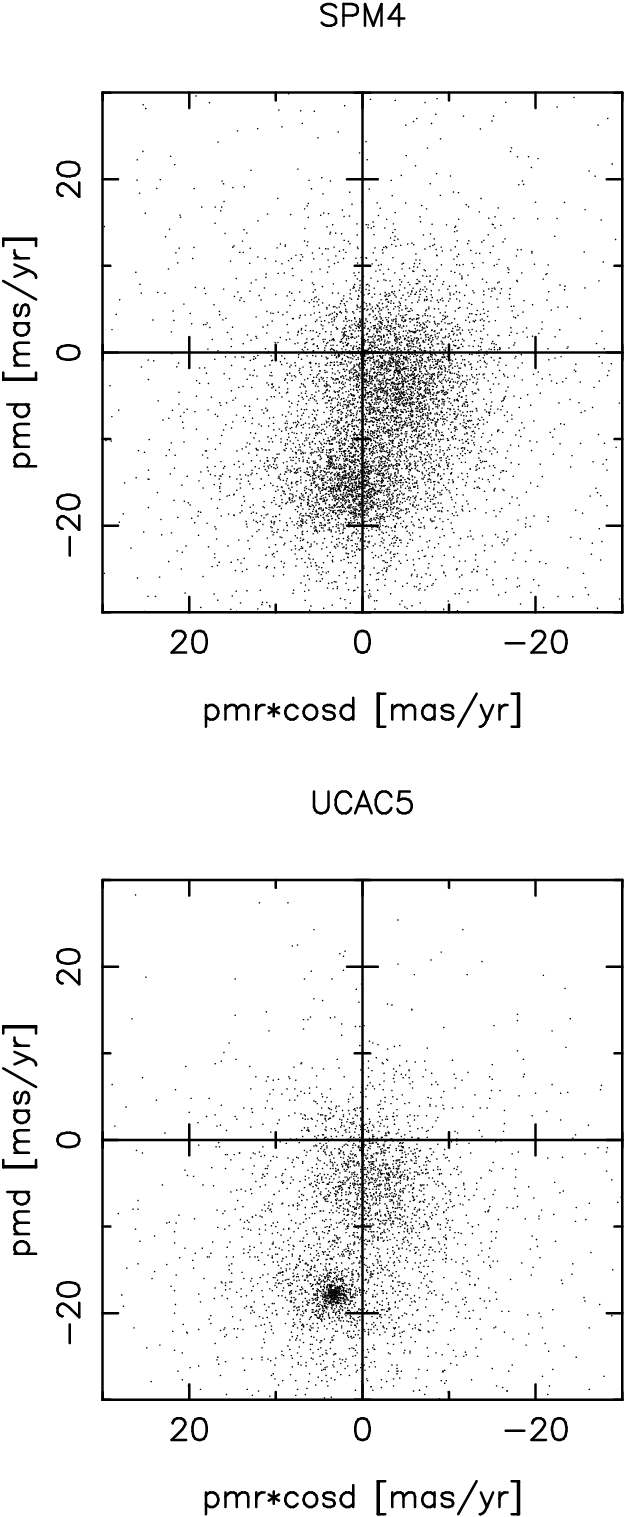}
\caption{SPM4 (top) and UCAC5 (bottom) proper motions 
(pmr = along RA, pmd = along Dec) of stars in common between UCAC5 
and SPM4 in the 30 arcmin area around the globular cluster NGC 6397.}
\end{figure}






\clearpage


\begin{table}
\begin{center}
\caption{Summary of relevant UCAC data.}
\begin{verbatim}
astrograph aperture  =    208.0 mm
focal length  (f/10) =   2060.0 mm
image scale          =    100.5 arcsec/mm
fixed bandpass       =  579 - 643 nm
field of view (lens) =  9 deg diameter

number of pixels CCD =  4k by 4k (Kodak front illum.)
pixel size           =  9.0 micrometer
pixel scale          =  0.905 arcsec/px
field of view CCD    =  1.02 by 1.02 deg
typical FWHM images  =  1.7 to 2.5 px

observing at CTIO    =  1997-2001 (Dec = -90 to +25)
observing at NOFS    =  2001-2004 (Dec = +25 to +90)
survey pattern       =  2-fold, center-corner
long  exposures      =  150 or 125 or 100 sec
short exposures      =  1/5 of long exposure

total number of exposures taken = 274,000
number of acceptable exposures  = 218,000

\end{verbatim}
\end{center}
\end{table}



\begin{table}
\begin{center}
\caption{Data columns of UCAC5 catalog.}
\begin{tabular}{rlcl}
\tableline\tableline
column &  name & unit &   description \\
\tableline
  1 &  srcid &           & Gaia source ID \\
  2 &  flg   &           & 1 = TGAS,  2 = not TGAS, in NOMAD, 3 = neither\\
  3 &  nu    &           & number of images used for mean position \\
  4 &  epoc  & 1/1000 yr & mean UCAC epoch (after 1997.0) \\
  5 &  ira   & mas       & mean UCAC RA  at epoch (item 4)  \\
  6 &  idc   & mas       & mean UCAC Dec at epoch (item 4) \\
  7 &  pmra  & 0.1 mas/yr& proper motion RA*cosDec  \\
  8 &  pmdc  & 0.1 mas/yr& proper motion Dec      \\
  9 &  pmer  & 0.1 mas/yr& formal error of proper motion RA*cosDec \\
 10 &  pmed  & 0.1 mas/yr& formal error of proper motion Dec \\
 11 &  gmag  & mmag      & Gaia DR1 G magnitude    \\
 12 &  umag  & mmag      & mean UCAC model magnitude \\
 13 &  Rmag  & mmag      & photographic R magnitude from NOMAD \\
 14 &  Jmag  & mmag      & 2MASS J magnitude  \\
 15 &  Hmag  & mmag      & 2MASS H magnitude  \\
 16 &  Kmag  & mmag      & 2MASS K magnitude  \\
\tableline
\end{tabular}
\end{center}
\end{table}


\clearpage


\begin{table}
\begin{center}
\caption{UCAC5 proper motions of extragalactic soruces from LQAC3.}
\begin{tabular}{ccccrrrrrrrrr}
\tableline\tableline
  nu      &  obj.    &  mag        &  z    & total  &
\multicolumn{2}{r}{mean [mas]} & 
\multicolumn{2}{r}{ RMS [mas]} & 
\multicolumn{2}{r}{norm. mean} & 
\multicolumn{2}{r}{norm. RMS}  \\
  $\ge$ & type &  $\le$ &  $\ge$ & n.obj. & 
     $\mu_{\alpha}$ &     $\mu_{\delta}$ & 
     $\mu_{\alpha}$ &     $\mu_{\delta}$ &   
     $\mu_{\alpha}$ &     $\mu_{\delta}$ & 
     $\mu_{\alpha}$ &     $\mu_{\delta}$ \\  
 \tableline
  2 &  all  &  all  &  all &  1108 &  $-$0.42 & $-$0.70 & 4.59 & 4.53 & 
                                      $-$0.11 & $-$0.20 & 1.14 & 1.11 \\
  2 &  R,Q  &  all  &  all &   461 &     0.05 & $-$0.39 & 3.76 & 4.20 & 
                                         0.02 & $-$0.12 & 0.87 & 0.99 \\
  2 &  all  &  15.8 &  all &   541 &  $-$0.99 & $-$0.99 & 4.88 & 4.34 & 
                                      $-$0.32 & $-$0.33 & 1.58 & 1.48 \\
  2 &  R,Q  &  15.8 &  all &   167 &  $-$0.83 & $-$0.31 & 3.74 & 3.87 & 
                                      $-$0.24 & $-$0.13 & 1.36 & 1.42 \\
  2 &  R,Q  &   all &  0.5 &   180 &  $-$0.21 & $-$0.08 & 4.06 & 4.50 & 
                                      $-$0.04 & $-$0.06 & 0.94 & 1.00 \\
  2 &  all  &   all &  0.5 &   184 &  $-$0.22 & $-$0.09 & 4.14 & 4.50 & 
                                      $-$0.03 & $-$0.07 & 0.96 & 1.00 \\
\tableline
  1 & ICRF  &   all &  all &   184 &     0.17 &    0.37 & 4.64 & 4.45 & 
                                         0.05 & $-$0.01 & 0.81 & 0.81 \\
  2 & ICRF  &   all &  all &   124 &     0.01 &    0.28 & 3.20 & 3.39 & 
                                         0.02 & $-$0.03 & 0.86 & 0.86 \\
\tableline
\end{tabular}
\end{center}
\end{table}


\end{document}